\documentclass[twocolumn,preprintnumbers,superscriptaddress,nofootinbib]{
revtex4}
\usepackage{graphicx}
\usepackage{amssymb}
\usepackage{color}
\usepackage{enumerate}
\def\beq{\begin{equation}}
\def\eeq{\end{equation}}
\def\beqn{\begin{eqnarray}}
\def\eeqn{\end{eqnarray}}

\renewcommand{\texttt}{{}}
\newcommand{\be}{\begin{eqnarray}}
\newcommand{\ee}{\end{eqnarray}}

\begin{document}

\title{Inflation in (Super-)renormalizable Gravity}

\author{Fabio Briscese}
\affiliation{Istituto Nazionale di Alta Matematica Francesco
Severi, Gruppo Nazionale di Fisica Matematica,
Citt\`a Universitaria, P.le A. Moro 5, 00185 Rome, EU}
\affiliation{SBAI, Sezione di Matematica, Sapienza Universit\`a di Roma,
Via Antonio Scarpa 16,  00161 Rome, EU}

\author{Antonino Marcian\`o}
%\email{amarcian@haverford.edu}
\affiliation{Department of Physics \& Astronomy, HB 6127
     Wilder Lab, Dartmouth College, Hanover, NH 03755, USA}
%\affiliation{Department of Physics and Astronomy, Haverford College,
%Haverford, PA 19041, USA}
%\affiliation{Department of Physics, Princeton University, New Jersey
%08544, USA}

\author{Leonardo Modesto}
%\email{lmodesto@perimeterinstitute.ca}
\affiliation{%Perimeter Institute for Theoretical Physics, 31 Caroline St.,
%Waterloo, ON N2L 2Y5, Canada.}
Department of Physics \& Center for Field Theory and Particle Physics,
Fudan University, 200433 Shanghai, China}

\author{Emmanuel N.~Saridakis}
%\email{Emmanuel$_-$Saridakis@baylor.edu}
\affiliation{Physics Division, National Technical University of Athens,
15780 Zografou Campus, Athens, EU}
\affiliation{Instituto de
F\'{\i}sica, Pontificia Universidad Cat\'olica de Valpara\'{\i}so,
Casilla 4950, Valpara\'{\i}so, Chile}

%\date{\small\today}

\begin{abstract}
\noindent
We investigate a (super-)renormalizable and ghost-free theory of gravity,
showing that under a natural (exponential) ansatz of the form factor and a
suitable truncation it
can give rise to the Starobinsky inflationary theory in cosmological
frameworks, and thus offering a theoretical justification of its origin.
We study the corresponding inflationary evolution and we examine the
generation of curvature perturbations, adapting the $f(R)$-like equations
in a symmetry-reduced FLRW metric. Furthermore, we analyze how the
ultraviolet regime of a simply renormalizable and unitary theory of gravity
is also compatible with the Starobinsky action, and hence we show that such
a theory
could account for an inflationary phase of the Universe in the
ultraviolet regime.
\end{abstract}

\pacs{05.45.Df,98.80.-k,  04.60.Pp}
\keywords{perturbative quantum gravity, nonlocal field theory, inflation}

\maketitle

\section{Introduction}

\noindent Recently a new gravitational action principle has
been introduced and/or reconsidered
in order to alleviate the shortcomings of Einstein
theory  \cite{BM,modesto1,modesto2, modesto3,modesto4}.
 This theory, and the corresponding approach to quantum gravity
partly inspired by the
Cornish and Moffat papers \cite{Moffat}, is
constructed in order to fulfill a synthesis of minimal hypotheses: (i)
classical solutions must be singularity-free; (ii) Einstein-Hilbert action
should be a good approximation of the theory at a much smaller energy scale
than the Planck mass;  (iii) the spacetime dimension has to decrease with
the energy in order to have a complete quantum gravity theory in the
ultraviolet regime; (iv) the theory has to be perturbatively renormalizable
at the quantum level (this requirement is strongly related to the previous
one), (v) the theory has to be unitary, with no other pole in the
propagator in addition to the graviton, (vi) spacetime is a single
continuum of space and time and in particular the Lorentz invariance is not
broken, consistently with observations.  This (super-)renormalizable
gravitational theory is therefore   consistent with the basic requirements
of quantum gravity.

On the other hand, a large amount of research has been dedicated
to the construction of modified gravitational theories, which are
capable of describing the observed late-time acceleration and the
detected signatures of inflation (see \cite{Nojiri:2006ri} and
references therein). Amongst them the $f(R)$-gravity (see
\cite{DeFelice:2010aj} and references therein) is perhaps the most
investigated one, with interesting implications in spite of its
simple form. However, although the above modified-gravity
scenarios exhibit interesting phenomenology in agreement with
observations, their construction is mainly artificial, without a
theoretical justification, hoping that an underlying fundamental
theory of quantum gravity, unknown up to now, would eventually
provide them as low-energy limits. In the same lines,
gravitational modifications based on the inclusion of finite
higher derivatives (see \cite{Clifton:2011jh} and references
therein) can also have interesting cosmological implications. For
instance, non local theories with lagrangian density $\mathcal{L}
= F(R, \Box R, \Box^2 R,\ldots,\Box^m R, \Box^{-}
R,\ldots,\Box^{-m} R)$ have a classical equivalent scalar
representation obtained introducing $2(m+n)$ auxiliary scalar
fields, which is capable to explain inflation and dark energy in a
unified framework \cite{odintsov1,odintsov3}, see also
\cite{odintsov2} for a complete review. However, all these
extensions of general relativity seems to be artificial too, with
the main exception being the Horava-Lifshitz gravity, in which the
higher-order derivatives are added following the fundamental
symmetries of the theory \cite{horava} and the inclusion of
fermionic matter has been conjectured to be responsible for the
perturbatively consistent UV behavior \cite{AMM}. See also 
\cite{Deser:2007jk} for an investigation of non-locally modified models of gravity, compatible with quantum loop corrections, in light of theory application as a mechanism for current cosmic acceleration.

In the present work we are interested in providing a theoretical
justification of some of $f(R)$-gravitational scenarios, in the
context of the above (super-)renormalizable gravitational theory.
In particular, we show that this theory, which in principle contains
infinite number of higher derivative terms, under a suitable truncation can
give rise to the
famous Starobinsky theory \cite{Staro} \be
\label{starobinskylagrangian}
\mathcal{L} = R + \epsilon R^2,
 \ee
which proves to be a viable inflationary model in perfect agreement with
current observations. Especially the recently released WMAP nine-year
results suggest that Starobinsky model is the one that best describes the
data, comparing to scalar field inflation \cite{wmap9}. Starobinsky model
was first introduced
due to one loop contributions of
conformally covariant matter fields to the Einsteinian theory of gravity
\cite{Staro}, but it was subsequently shown that it can
be derived in the framework of superstring theory
\cite{bailin,Mijic,dolgov}, or that it  can be embedded in M-theory
\cite{odintsovMtheory} and in $F(\textsl{R})$ supergravity
\cite{starobisnkyembedding}. In the following we show that the Starobinsky
model can alternatively and naturally emerge from a consistent truncation
of the
aforementioned (super-)renormalizable quantum gravity models, when a proper
form-factor is assumed.

The plan of the work is as follows: In section \ref{SRG} we review
(super-)renormalizable gravity and we show how the Starobinsky model can
arise under a suitable {\it ansatz} for the form factor. In section
\ref{sectioninflationSRG} we examine the inflation realization in such a
scenario and we provide the corresponding values of observables. Finally,
in section \ref{Discussion} we discuss on the obtained results, while
section \ref{Conclusions} is devoted to the conclusions.

In what follows we
use the signature $(+ - \dots -)$, and the
curvature tensor is defined as $R^{\alpha}_{\beta \gamma \delta} =
- \partial_{\delta} \Gamma^{\alpha}_{\beta \gamma} + \dots$, the
Ricci tensor as $R_{\mu \nu} = R^{\alpha}_{\mu \nu \alpha}$, and
the curvature scalar as $R = g^{\mu \nu} R_{\mu \nu}$, where
$g_{\mu \nu}$ is the metric tensor \cite{staro2}.

\section{(Super-)renormalizable gravity}
\label{SRG}

\noindent Let us briefly review   super-renormalizable gravity
(SRG). In order to be more transparent, we first describe the
general theory and then we provide the Starobinsky theory as a
special case.

\subsection{General theory}

\noindent Super-renormalizable gravity (SRG) is well defined
perturbatively at the quantum level. Additionally, at the
classical level, the gravitational potential \cite{BM}, the black
hole solutions \cite{NS,modesto1, modesto2,
modesto3,ModestoMoffatNico} and the cosmological solutions are
singularity-free \cite{BM,Bis2}. The
corresponding gravitational Lagrangian is a ``non-polynomial''
extension of the renormalizable quadratic Stelle theory
\cite{Stelle} and it has the following general structure: \be \!
\mathcal{L} =  R - R_{\mu \nu} \gamma(\Box_{\Lambda}) R^{\mu \nu}
+ \frac{1}{2} R \,\gamma( \Box_{\Lambda})  R \, ,
\label{theory}
\ee
where the ``form factor'' $\gamma(\Box_{\Lambda})$ is an
``entire function'' of the covariant D'Alembertian operator and
$\Box_{\Lambda} = \Box/\Lambda^2$, with $\Lambda$ is an invariant
mass scale. We mention that non-locality only involves positive
powers of the D'Alembertian operator since the two form factors
are entire functions. The above theory is not unique, but all the
freedom present in the action can be embedded in the function
$\gamma(\Box_{\Lambda})$ \cite{efimov,Tombo}. Such function must be
interpreted in analogy with the interaction of a photon with a
nucleon, that is the form factor for gravity $\gamma(
\Box_{\Lambda})$ could be eventually measured experimentally.

It proves convenient to express the form factor
$\gamma(\Box_{\Lambda})$ introducing a new form factor
$V(\Box_{\Lambda})$ that appears in both the spin two and spin
zero part of the propagator, namely \be \gamma(\Box_{\Lambda})
\equiv \frac{V(\Box_{\Lambda})^{-1} -1}{\Box}\,. \label{FFS}
 \ee
The above choice is essential in order to have a unitarity ghost-free
theory \cite{modesto1, modesto2, modesto3, BM}, and the classical
Lagrangian simplifies to

\be \hspace{-0.3cm} \mathcal{L} =  R - G^{\mu \nu}
\left(\frac{V(\Box_{\Lambda})^{-1} -1}{\Box} \right)R_{\mu\nu} \,
,
 \label{compact}
\ee
where $G^{\mu \nu}$ is the Einstein's
tensor. In order to better clarify this point we recall here the
gauge invariant two-point function for the theory
(\ref{compact}):
\be [\mathcal{O}^{-1}(k)]_{\rm gauge\,\, inv.} =
\frac{V(k^2/\Lambda^2)  } {k^2} \left( P^{(2)} - \frac{P^{(0)}}{2
}  \right)\,, \label{propgauge2} \ee where $\mathcal{O}(k)$ is the
kinetic operator arising from an expansion of the gravitational
action around the flat metric $\eta_{\mu \nu}$ in powers of the
graviton field $h_{\mu \nu}$, defined by $g_{\mu \nu} = \eta_{\mu
\nu} +  h_{\mu \nu}$. Since $V(\Box_{\Lambda})$ in (\ref{FFS}) is
a non-polynomial trascendental entire function, only a massless
graviton propagates and thus the theory is ghost free. Moreover,
the above theory is super-renormalizable \cite{modesto1, modesto2,
modesto3}, as well as unitary and microcausal \cite{efimov}.

In this paper we will mainly focus on the following very natural form
factor
$\label{formfactor1} V(\Box_{\Lambda}) = e^{ - \Box_{\Lambda}}$.
This ansatz corresponds to a specific and notable super-renormalizable
theory, since in \cite{modesto2} it was shown that in this case the
graviton propagator is the same with the one obtained starting from a
theory of gravity endowed with $\theta$-Poincar\'e quantum groups of
symmetry.
Other possible interpretations of such choice come from string filed theory
\cite{collective1,collective2,BisSF,Wnc,noArchimede},
stochastic fluctuations of the spacetime at short distance \cite{Namsrai}
or if we intrinsically assume
fractal properties of the spacetime when it is probed at high energy
\cite{CF1,CF2,CF3,CF4, CF5}.
Finally, it would be intriguing to consider the role of the cut-off in defining a minimal length, following the perspective of possible phenomenological  explorations (see for instance Ref.~\cite{SpreBleiNi}).

\subsection{Embedding the Starobinsky $R+\epsilon R^2$ model in
super-renormalizable
gravity}\label{srg}

\noindent
At the classical level  we can truncate the theory (\ref{compact})
at will. In this subsection we focus on the first correction to
the Einstein Hilbert action.  Since we are dealing with a $D=4$
Friedmann-Lema\^itre-Robertson-Walker (FLRW) metric, the following term
turns
out to be topological
\be
 \int d^4 x \sqrt{|g|} \left[3 R_{\mu
\nu} R^{\mu \nu} - R^2 \right] = {\rm topological} \, ,
\ee
 which
reduces the truncated theory to
 \be
 \mathcal{L} =  R + \frac{1}{6
\Lambda^2} R^2 + O\left( \frac{R \, \Box R}{ \Lambda^4} \right).
\label{TRU}
\ee
When $R \Box R/\Lambda^2 \ll R^2$ the above Lagrangian
reduces to the Starobinsky inflationary model $R+\epsilon R^2$,
that is the super-renormalizable gravity offers an alternative
explanation from its origin at the fundamental level.

In
order to obtain a realistic cosmological application of the above
model, one needs to include the matter Lagrangian $\mathcal{L}_M$,
corresponding to an energy-momentum tensor $T_{\mu \nu}$. Then the
total action in a universe governed by the above truncated
super-renormalizable gravity writes as
 \be S = \frac{1}{2 \kappa^2}
\int d^4 x \sqrt{|g|} \left[R + \frac{1}{6 \Lambda^2} R^2 +
\mathcal{L}_{M} \right]\,,
 \label{fullaction}
 \ee
 where $\kappa^2
= 8 \pi G_N$ and $G_N$ is the gravitational constant.

We conclude that the action  (\ref{compact}) can be
truncated and then recast in the language of $f(R)$ theories,
provided that $f(R)=R+R^2/(6 \Lambda^2)$, which is the famous
Starobinsky model \cite{Staro,Mijic}. The resulting truncated
equations of motion are recovered to be
\be \label{FRtrunca}
&& R_{\mu \nu} - \frac{1}{2} g_{\mu \nu} R
+ \frac{R}{3 \Lambda^2}   \left( R_{\mu \nu} - \frac{1}{4} g_{\mu \nu} R
\right)  \nonumber \\
&&
- \frac{1}{3 \Lambda^2} \left( g_{\mu \nu}
\Box R - \nabla_{\mu} \nabla_{\nu} R \right) =  8 \pi G_N T_{\mu \nu}.\ \ \
\ee
It is interesting to mention that starting from an infinite number of
higher-derivative terms and under the above suitable truncation,
super-renormalizable gravity gives rise to a specific $f(R)$ gravity
model, which in some sense is a unification of the different classes
in general modified-gravity, {\it i.e.} the $f(R)$ and the explicitly
higher-derivative one.

\subsection{Embedding the Starobinsky $R+\epsilon R^2$ model in
renormalizable gravity} \label{section RG}

\noindent
For completeness, in this subsection, we explore another specific
choice of the form factor $V(z)$, which introduces the construction of
renormalizable gravity (RG). In particular, choosing
\begin{equation}
\label{modefactor}
 V(z)^{-1} = e^{H( z )},
\end{equation}
 with $z : = - \Box_{\Lambda}$ and
\begin{equation}
 H(z) = \frac{1}{2} \left[ \gamma_E + \Gamma \left(0,
z^{2}\right) \right] + \log | z |  \, , \,\,\,\, {\rm Re}( \,
z^{2} ) > 0 \nonumber \,,
\end{equation}
the ultraviolet limit of the form
factor is \be \lim_{z \rightarrow + \infty} V(z)^{-1}  = - z \, .
\label{limit} \ee
Since the entire function $\gamma(z)$ approaches a constant for
$|z| \rightarrow +\infty$, this theory embodies the quadratic
Stelle action in the ultraviolet limit, but without any ghost
pole in the propagator. The form factor cross-connects the quadratic
action in the infrared, with an equivalent theory in the ultraviolet.
The amplitudes are divergent at each order in the loop expansion
and the maximal superficial degree of divergence is four, similarly to
the local Stelle's theory. Thus, the theory ceases to be
super-renormalizable, but it preserves  renormalizability
and unitarity as it can be inferred from the general structure
of the propagator (\ref{propgauge2}) within the form factor
(\ref{modefactor}).

Let us consider now the intermedium regime. We can write the
entire function $H(z)$ as a series
\be
 H(z) = \sum_{n =1}^{+
\infty} ( -1 )^{n-1} \, \frac{z^{2 n}}{2n \, n!} = \frac{ z^{
2}}{2} - \frac{ z^{ 4}}{8} + O( z^6)\,.\label{HS}
 \ee
The theory for an FLRW spacetime coincides with the Starobinsky
theory in the high energy regime, and it is well approximated by
the same theory at lower energies too. In particular, the
ultraviolet Lagrangian is exactly \be \label{RGlagrangian}
\mathcal{L}_{\rm UV} \approx R + \frac{1}{6 \Lambda^2} R^2\,  .
\ee
 Applying the Starobinsky and/or Mijic \cite{Staro, Mijic}
analysis yields a general class of solutions enjoying initial
quasi-de Sitter inflationary behavior. Nevertheless, we are still
confronted with the issue of connecting the inflationary epoch to
the current FLRW universe phase of expansion. Relation
(\ref{RGlagrangian}) is indeed valid at high curvatures and, as
long as $R$ decreases during expansion, higher derivative
corrections to (\ref{RGlagrangian}) may be important, at least for
a short period, and may force the exact solutions to be different.
Thus, we cannot easily conclude that the theory (\ref{limit})
gives a suitable cosmological expansion, since we should
investigate carefully the behavior of the solution in the
intermedium regime (\ref{HS}), the reheating and the connection
with FLRW universe. Therefore, even though one expects that higher
derivative terms are negligible, a deeper analysis of the
cosmological solutions in the intermedium regime is needed in
order to exclude classical instabilities. We leave such a detailed
investigation for a future work.

\subsection{Differences with string
theory} \label{differences}

\noindent
We finish this section by referring to  some differences with the string
theory scenarios. In particular, in superstring theory the
effective action of the point-particle limit of the
ten-dimensional Lagrangian is made out of the usual
Einstein-Hilbert term plus the   operators
 \be
 R_{\mu \nu
\lambda \rho} R^{\mu \nu \lambda \rho} + a R_{\mu \nu} R^{\mu \nu}
+ b R^2\,, \nonumber
\ee
where $a$ and $b$ are constants. After
compactification the four dimensional theory reads as
\be
\label{stridue}
 \mathcal{L}_{\rm string} = R + \left(
\frac{a+1}{3} + b \right) \frac{G_N {\rm V}_6}{\langle \phi
\rangle} R^2\,,
\ee
 where ${\rm V}_6$ is the compactified volume
of the six ``extra" dimensions  and $\langle \phi \rangle$ is the
expectation value of the scalar dilaton field. Therefore, using arguments
and taking inspiration from string theory, we might think of
fixing uniquely the coefficient in front of $R^2$ in the SRG and
RG.
However, in string theory the preferred values for $a,b$ are
$a=-4$ and $b=1$ \cite{Mijic,maggiore}, and thus comparison of 
(\ref{stridue}) and
(\ref{starobinskylagrangian}) would  suggest that $\epsilon
= 0$,
namely no $R^2$ term is favored by superstring theory.
However, these values for  the parameter $a$ and $b$ are ambiguous because
sensible to a metric field
redefinition \cite{W11}. %
On the other
hand, in
(super-)renormalizable gravity, if a Starobinsky limit of the
theory exists, the corresponding coefficient is uniquely fixed. In
addition, in
such a theory any form factor of the form $V(\Box_{\Lambda})^{-1} \approx 1
+
\Box_{\Lambda}+ O(\Box_{\Lambda}^2)$ is compatible with the Starobinsky
inflationary model. Thus, we can see that Starobinsky scenario can arise
in SRG and RG more effectively than in string theory.

\section{Inflation in super-renormalizable
gravity}\label{sectioninflationSRG}

\noindent
In the previous section we saw that the
super-renormalizable theory, when $R \Box
R/\Lambda^2 \ll R^2$, reduces to the Starobinsky inflationary
model $R + \epsilon R^2$ with $\epsilon = 1/6\Lambda^2$. Thus,
provided that such an approximation starts to be valid at the
beginning of inflation, and hence lasts up to the present epoch,
in the proposed SRG model one can obtain the same inflationary
picture as in the Starobinsky theory. In the following we
summarize the main properties of SRG  model introduced in section
\ref{srg}, as they are deduced by $R + \epsilon R^2$ Starobinsky model.

In \cite{Staro} it has been shown that the Starobinsky theory
admits an unstable de Sitter phase, and in \cite{Mijic} the
authors have shown that the same theory entails a general class of
solutions describing with a certain accuracy the cosmological
evolution of the Universe. Indeed, such solutions behave initially as a
quasi-de Sitter universe, with a slowly decreasing Hubble parameter
\be \label{Hsol} H(t) \simeq H_0 - \frac{\Lambda^2}{6} (t-t_0)\,,
\ee
with $H_0$ the Hubble parameter at the beginning of inflation,
namely at time $t_0$. Having lasted for a time $t_{osc}-t_0 \simeq
6 H_0/\Lambda^2$, inflation ends and the Universe enters a phase
of oscillations with
\be &&H(t) \simeq f(t) \cos^2(\omega t)\,,
\ee
where $\omega \simeq
\Lambda/2$ and with
\be
 &&f(t) \simeq 1/\left[3/\omega + 3(t-t_{osc})\right]
\ee
representing the damping factor for the amplitude's oscillations.
During the oscillatory phase the Universe is reheated and
standard-model particles are produced with a reheating temperature \cite{Mijic}
\be
 T_r \simeq 3 \times 10^{-2} \, \epsilon^{-\frac{1}{2}}
 \simeq 4 \times 10^{17}  {\rm GeV}\, 
\sqrt{6\Lambda^2/ M_p^2}\,,
\ee
where $M_p$ denotes the Planck mass.

It should be noticed that one needs $T_r \simeq
10^{10}\!-\!10^{16}$ GeV in order to avoid
the monopole problem and properly account for baryogenesis.
Finally, after reheating, the Universe enters the usual FLRW phase
and then evolves as in the standard cosmological
picture\footnote{A cosmological constant is still needed in order to
describe the late-time acceleration.}.

The $R + \epsilon R^2$ scenario entails a spectral index $n_s -  1 \simeq - 2/N_e \simeq -0.04 \times (50/N_e)$  and a tensor-to-scalar ratio $r \simeq 12/N_e^2 \simeq 0.005 \times (50/N_e)^2$, parameterized in terms of the e-foldings number $N_e$ which is $\Lambda$-independent within a good approximation \cite{starobisnkyembedding}. Agreement with WMAP nine-year  data \cite{wmap9}, accounting for $n_s = 0.971\pm0.010$ and $r \!<\! 0.13$ at $95\%$ confidence level, is recovered provided that $N_e \simeq 50-55$ \cite{starobisnkyembedding,wmap9}. Moreover the amplitude of initial perturbations requires \cite{starobisnkyembedding}, 
\be \label{coscon}
\frac{\Lambda}{M_p} \simeq 1.5 \times 10^{-5}\times (50/N_e)\,.
\ee

Taking into account the constraints from cosmological perturbations in (\ref{coscon}), the reheating temperature turns out to be $T_r\simeq 5\times 10^{12}$ GeV. Assuming supersymmetry, a potential issue related to such a value of $T_r$ is the overproduction of gravitinos, which would take place at reheating temperatures $T_r \geq 10^{10}$ GeV for a wide range of gravitino's masses \cite{Kawasaki:1994af, Moroi:1995fs}. The role of a $R^3$ term in lowering the reheating temperature down to $T_r \simeq 10^{9}$ GeV  has been addressed in \cite{starobisnkyembedding}. Such a term also originates within the framework of the higher-dimensional extension of the SRG theories we have addressed in this paper. We leave to a forthcoming paper\footnote{We acknowledge the referee for pointing out to us the issue of gravitinos over-production.} the task of establishing a quantitative relation with the study developed in \cite{starobisnkyembedding}.

In light of these considerations, the SRG model constructed in
subsection \ref{formfactor1} provides a suitable inflationary
scenario which mimics very well the Starobinsky theory, to which
it reduces at low curvatures $R \Box\, R / \Lambda^2 \ll R^2$.
Additionally, the SRG model can be related  to FLRW universe at
late times. However, as we discussed in subsection \ref{section RG},
this is not the case for the RG model introduced there, since even
though the latter model reduces asymptotically to $R+\epsilon R^2$
gravity for $R \rightarrow \infty$,  it may deviate from the
Starobinsky theory as long as the curvature $R$ decreases, and
thus the relation between the RG model and the FLRW universe will
be non-trivial in this case. Finally, we  remark that the mass
scale $\Lambda$, which represents a cutoff for the higher
derivative terms, is well below the Planck scale, and thus it may
have observational consequences for highly energetic but still
sub-Planckian phenomena.

\section{Discussion}
\label{Discussion}

\noindent
Let us make some remarks concerning the equivalence of our particular SRG
model of Lagrangian  (\ref{TRU})
with the $f(R)=R+\epsilon R^2$ gravity. As we analyzed in subsection
\ref{srg}
the Starobinsky model  can be obtained through a
coherent truncation of SRG, and thus in section
\ref{sectioninflationSRG} we used this result in order to describe the
inflationary scenario in SRG. However, some significant
difference could arise between SRG and Starobinsky model.

As it is well known, a general property of $f(R)$ gravity is the
appearance of
an extra scalar curvaton degree of freedom. Such a scalar degree
of freedom can be explicitly expressed performing a Weyl
transformation of the metric tensor $ \tilde{g}_{\mu \nu}(x)
\equiv \exp\left[\sqrt{2/3} \, k \, \phi\right]\, g_{\mu \nu}(x) $,
which maps the original $f(R)$ theory of the Jordan frame into General
Relativity plus a   scalar field in the Einstein frame, provided that
the scalar field potential is $ V(\phi)= \left[f'(R) \tilde{R} -
f(R)\right]/2 k^2 f'(R)^2$, where $f'(R) \equiv \partial
f(R)/\partial R$ and $\tilde{R}$ is the Ricci scalar constructed
with the $\tilde{g}_{\mu\nu}$ metric tensor (for instance see
\cite{odintsov}). Therefore,  the phenomenology of the $f(R)$
scenario could be obtained in terms of the scalar field $\phi$ in the
Einstein frame. For example the number of e-folds during inflation is
recovered to be $N_{e}= \int_t^{t_f}\,H dt \simeq  \int_{\phi_{f}}^{\phi}
V/V' M_p^2 d\phi $, where $t_f$ denotes the time at the end of
inflation.

What is more important in the above picture is that the extra scalar
degree of
freedom is responsible for scalar perturbations during inflation. Thus, in
the Einstein frame the slow-roll inflation parameters are
obtained in terms of the scalar field $\phi$   as $\epsilon_\phi=
\left(M_p^2/2\right)  \left(V'/V \right)^2$ and  $ \eta_\phi =
M_p^2  V''/V$  \cite{LL}, which lead to the slope of the power
spectrum: $ n_s=1 + 2 \eta_\phi - 6 \epsilon_\phi $. Finally,
the amplitude of initial perturbations is given by $\Delta^2_R =
M_p^4 V/(24 \pi^2 \, \epsilon_\phi)$, which is observationally
estimated \cite{LL} to be $ \left(V(\phi)/\epsilon_\phi
\right)^{\frac{1}{4}} \simeq 6.6 \times 10^{16}\, {\rm GeV} $.

Therefore, as we analyzed in detail in the previous section,
in the specific case of $f(R)=R + \epsilon R^2$ gravity the above analysis
allows to
estimate the $\Lambda$ parameter as $\Lambda/M_p \sim 10^{-5}$
\cite{starobisnkyembedding,ScalarR2}, and correspondingly for the   power
spectrum $n_s = 1 + 2 \eta -6\epsilon \simeq
1-2/N_e$, for the tensor primordial spectrum $n_t  \simeq -2 \epsilon
\simeq - 3/2 N_e^2$, and for the tensor-to-scalar  ratio $ r= 16
\epsilon\simeq 12/N_e^2$ \cite{ScalarR2}. In other words,
 in the $f(R)=R+\epsilon R^2$ model the induced extra scalar
degree of freedom $\phi$ is responsible for the generation of
primordial scalar perturbations  which in turn are used to
estimate the $\epsilon$ parameter, or the $\Lambda$ parameter in
the SRG case.

However, a general feature of SRG and RG  which makes a
fundamental difference with $R+\epsilon R^2$ theory, is the fact
that no extra scalar degree of freedom seems to exist in such
theories. In fact, in  \cite{modesto1, modesto2, modesto3, BM}
it has been shown that, after quantization of the gravitational
field on flat Minkowskian background metric, only the usual
massless spin-two gravitons and no other extra degrees of freedom
are found. We stress that this result has been demonstrated only
on a flat background and it is not trivially extended to a curved
spacetime. However, if confirmed, this would in part invalidate
the analysis performed in subsection \ref{sectioninflationSRG}. In
particular, at a classical background level the evolutionary picture
described in subsection \ref{sectioninflationSRG} is safely valid,
since the truncation introduced in subsection \ref{srg} is valid as
long as $R \Box R/\Lambda^2 \ll R^2$. However, at perturbation
level the absence of the extra degree of freedom (if confirmed) could
leave the scenario without a mechanism to generate primordial perturbations
during the inflationary epoch, and thus one should need to find other such
mechanisms, for instance introducing  by hand additional scalar or
alternatively vector fields
\cite{gaugeflation} (such a formulation would also imply that
the obtained bound on the parameter $\Lambda \sim 10^{-5} \times M_p$ would
need to change).

Nevertheless, note that SRG can have an advantage comparing to
$f(R)=R+\epsilon R^2$ gravity, concerning the generation of
non-gaussianities. In particular, it is known that the effective scalar
degree of freedom of $f(R)=R+\epsilon R^2$ gravity is not capable to
correctly produce the non-gaussianities.
On the other hand, SRG has still the choice to add higher order terms in
the Lagrangian (\ref{TRU}), which could lead to non-gaussian
perturbations. Since upper bounds on non-gausianities may be soon
imposed through the experimental investigation of CMBR  by the WMAP
\cite{creminelli} and the Planck satellites,
it will furnish an immediate opportunity to falsify models with large
non-gaussianities, in order to distinguish among the plethora of
inflationary models of the literature. We mention that the amount and shape
of deviations from a Gaussian distribution of primordial density reveals a
critical dependence on the details of the given inflationary model
\cite{Bart, Chen}, and thus it could be a crucial test for the SRG origin
of Starobinsky scenario.

\section{Conclusions}
\label{Conclusions}

\noindent
In this work we investigated the realization of inflation in the context
of (super-)renormalizable gravity, which is a gravitational theory
constructed consistently with the basic and minimal requirements of a
quantum gravity, being  well defined perturbatively at the quantum level.
The  corresponding gravitational Lagrangian is a ``non-local'' extension of
the renormalizable quadratic Stelle theory, expressed using a form factor
consisting of an entire function. Thus, at the classical level we can
truncate the theory in order to obtain many subclasses, one of them being
the Starobinsky inflationary model. Similarly, imposing suitably
a different choice for the form factor, instead of super-renormalizability
we can obtain a renormalizable theory that can also give rise to
Starobinsky model, though not so efficiently. In summary,
(super-)renormalizable gravity offers an explanation of the origin of
Starobinsky model at the fundamental level.

The approximate realization of the  Starobinsky model in
(super-)renormalizable gravity allows us to use the inflationary results
of that model in order to examine inflation in (super-)renormalizable
gravity. In particular, we find that the solutions of the theory
behave initially as a quasi-de Sitter universe, with a slowly decreasing
Hubble parameter, and thus inflation can end and the Universe
enter a phase of damping oscillations that reheats it.

However, there is still a difference between Starobinsky -type
inflation in $f(R)$ gravity and the Starobinsky-type inflation
in (super-)renormalizable gravity. In particular, while in the
former there appears an extra scalar curvaton degree of freedom
which is responsible for generating the perturbations during
inflation, in the latter case no extra scalar degree of freedom
seems to exist. Although this has not be proven for general
spacetimes but only for flat ones, if confirmed it would make
necessary an incorporation of a new mechanism to generate the
observationally required primordial perturbations. However,
inflation in (super-)renormalizable gravity has an advantage,
namely that it could lead to non-gaussian perturbations, which is
not the case in Starobinsky-type $f(R)$ gravity inflation. Such a
prospect would act as an additional asset for
(super-)renormalizable gravity.

\begin{acknowledgments}
\noindent
We warmly thank S. Alexander and S. Odintsov for valuable discussions and comments to the paper. F. Briscese is a Marie Curie fellow of the Istituto Nazionale di Alta
Matematica Francesco Severi. A. Marcian\`o gratefully acknowledges support
by Perimeter Institute during the early stage of this work and financial
support
through Career grant NSF (part of this work has been carried out while at
Princeton University). Research at Perimeter Institute is supported by the Government of Canada through Industry Canada and by the Province of Ontario through the Ministry of Research \& Innovation. 
The research of ENS is implemented within the framework of the Action
``Supporting Postdoctoral Researchers'' of the Operational Program
``Education and Lifelong Learning'' (Action's
Beneficiary: General Secretariat for Research and Technology), and is
co-financed by the European Social Fund (ESF) and the Greek State.
\end{acknowledgments}

\end{document}